\documentclass[12pt]{article}
\pdfoutput=1
\usepackage{graphicx}
\usepackage{amsmath,amsfonts,amssymb,stmaryrd,latexsym}
\usepackage{url,subcaption,comment}
\usepackage{hyperref}
\usepackage{xcolor}
\hypersetup{
    colorlinks=true,       
    linkcolor=purple,   
    citecolor=blue,        
    filecolor=magenta,      
    urlcolor=black           
}
\usepackage{cite}
\usepackage{dcolumn}
\usepackage{bm}

\renewcommand{\theequation}{\thesection.\arabic{equation}}
\newcommand{\be}{\begin{equation}}
\newcommand{\ee}{\end{equation}}
\newcommand{\bea}{\begin{eqnarray}}
\newcommand{\eea}{\end{eqnarray}}
\newcommand{\bear}{\begin{eqnarray}}
\newcommand{\eear}{\end{eqnarray}}
\newcommand{\ba}{\begin{array}}
\newcommand{\ea}{\end{array}}

\newcommand{\km}{\kappa}

\begin{document}

\baselineskip=18pt \pagestyle{plain} \setcounter{page}{1}

\vspace*{-1cm}


\begin{center}

{\large \bf    
Dijet and electroweak limits on a $Z'$ boson coupled to quarks
} \\ [9mm]

{\normalsize \bf Bogdan A. Dobrescu$^\diamond$, Felix Yu$^\star$ \\ [3mm]
{\small {\it $^\diamond$ Particle Theory Department, Fermilab,
    Batavia, IL 60510, USA }}\\
{\small {\it  $^\star$ PRISMA$^+$ Cluster of Excellence \& Mainz Institute for
  Theoretical Physics, Johannes Gutenberg University, 55099 Mainz, Germany}}\\
}

\center{December 10, 2021; Revised December 19, 2023}

\end{center}

\vspace*{0.2cm}

\begin{abstract}
An insightful way of presenting the LHC limits on dijet resonances is
the coupling-mass plot for a $Z'$ boson that has flavor-independent
quark interactions.  This also illustrates the comparison of low-mass
LHC sensitivity with constraints on the flavor-independent $Z'$ boson from electroweak
and quarkonium measurements.  To derive these constraints, we compute
the $Z'$ mixing with the $Z$, the photon, and the $\Upsilon$ meson,
emphasizing the logarithmic dependence on the masses of the new electroweak-charged fermions (``anomalons") required to cancel the gauge anomalies.  We update the
coupling-mass plot, extending it for $Z'$ masses from 5 GeV to 5 TeV.
\end{abstract}

\vspace*{0.7cm}

\renewcommand{\contentsname}{\normalsize\large Contents}
{\small
\hypersetup{linktocpage} 
\tableofcontents
\hypersetup{linkcolor=red} 
}

\newpage

\section{Introduction} \setcounter{equation}{0}

In the past few years, significant efforts have proven successful at
advancing hadron collider sensitivity to electroweak scale dijet
resonances.  At the end of Run 1 of the Large Hadron Collider (LHC),
in 2014, the ATLAS and CMS experiments had leading experimental
sensitivity to $\mathcal{O}(\text{TeV})$ dijet resonances, but
previous hadron collider experiments such as UA2 and CDF still
provided the leading constraint for resonances below a few hundred
GeV~\cite{Han:2010rf, Harris:2011bh, Dobrescu:2013coa, Chala:2015ama}.
The situation has now changed, with the advent of advanced triggering
techniques to overcome the intrinsic large quantum chromodynamic (QCD)
background at low dijet masses as well as dedicated efforts to probe
resonances in associated production modes~\cite{Shimmin:2016vlc,
  CMS:2016ltu, CMS:2017nuu, CMS:2017dcz, ATLAS:2018hbc, CMS:2018kcg,
  ATLAS:2018qto, ATLAS:2018tfk, ATLAS:2019itm, CMS:2019xai,
  CMS:2019emo, CMS:2019mcu, ATLAS:2020zzb}.  These more specialized
searches are complemented by the high-mass
analyses~\cite{ATLAS:2017eqx, CMS:2018mgb, CMS:2018wxx, ATLAS:2019fgd,
  CMS:2019gwf}, which have been impressively extended to dijet
resonances as heavy as several TeV.

Searches for dijet resonances are powerful probes of many theories
beyond the Standard Model (SM), because any particle produced in the
$s$-channel can decay back into two partons which then hadronize.  In
models with an additional $U(1)$ gauge symmetry, such as gauged baryon
number~\cite{Pais:1973mi, Carone:1994aa, Bailey:1994qv, Carone:1995pu,
  Aranda:1998fr, FileviezPerez:2011pt, Duerr:2013dza, Perez:2014qfa,
  Dobrescu:2014fca, Dobrescu:2015asa, Dobrescu:2017sue}, the
phenomenology of the associated $Z'$ boson is mainly characterized by
two parameters, the $Z'$ mass and its gauge coupling.  Searches
spanning different collider environments can then most easily be
interpreted in the coupling versus mass plane~\cite{Dobrescu:2013coa},
highlighting opportunities for further collider searches to cover
possible gaps in sensitivity.  

Here we reiterate that $Z'$ models generically include additional new
particles, and analyze how parameters associated with those particles
impact the $Z'$ properties.  The new particles include at least one
scalar associated with the $U(1)$ symmetry breaking sector, and some
fermions (``anomalons") charged under both the $U(1)$ and the SM gauge
groups, required to cancel the gauge anomalies.  Even when the $Z'$
boson cannot decay into non-SM particles, its mixing with the SM
spin-1 fields are impacted at 1-loop level by the masses and couplings
of the anomalons.

Nevertheless, the hadron collider limits are adequately captured by
the gauge coupling versus $Z'$ mass plot.  Comparing the limits from
hadron colliders with the electroweak data and other low-energy
constraints, however, needs a detailed analysis. We perform this
analysis and extend the coupling-versus-mass plot from 5 GeV to 5 TeV,
with exemplary choices of the anomalon parameters controlling the
mixing-induced constraints.  It turns out that the dependence on those
parameters affects only the low-energy constraints, and in a limited
fashion.

In Section~(\ref{sec:coupling-mass}), we provide an update of the
current status for weakly coupled, $q\bar{q}$, color-neutral vector
resonances and discuss associated phenomenology that can further the
experimental sensitivity in coming years.  After introducing a minimal
anomalon sector for gauged baryon number, in
Section~(\ref{sec:mixing}) we focus on the kinetic mixing operators
between the new $Z'_B$ boson and the $Z$ and $\gamma$ bosons of the SM
induced by the anomalon content.  The finite kinetic mixing effects
from the UV completion of gauged baryon number are also important for
the phenomenology of $Z'_B$ bosons lighter than the $Z$ boson.  We
reevaluate the constraints in the coupling--mass plane from mixing
with the $Z$ boson, $\Upsilon$ meson, direct $q\bar{q}$ resonance
limits from colliders, LEP limits on charged anomalon, and the
anomly-induced $Z \to Z'_B \gamma$ exotic decay in
Section~(\ref{sec:low-mass}).  We conclude in
Section~(\ref{sec:conclusions}), and a detailed discussion of our
kinetic mixing calculation is presented in
the Appendix.

\bigskip

\section{Dijet resonance limits in the coupling-mass plane}
\setcounter{equation}{0}
\label{sec:coupling-mass}

A color-singlet, electrically-neutral spin-1 particle, usually
referred to as a $Z'$ boson, may have renormalizable couplings to the
SM quarks.  As we are interested in bosons of a wide range of masses,
including at or below the electroweak scale, the simplest set of
couplings is flavor diagonal and universal, as described by the
following Lagrangian terms: 
\be
\frac{g_{_B}}{2} Z^\prime_{B\mu} \sum_q \, 
\left( \frac{1}{3}\, \overline q_L \gamma^\mu q_L + 
\frac{1}{3}\, \overline q_R \gamma^\mu q_R \right)  ~~.
\label{eq:quark-couplings}
\ee
The overall coupling, $g_{_B}$, is typically of order one or
smaller. Its normalization (the factor of 1/2) is chosen to be similar
to the SM $Z$ coupling (if the hypercharge coupling is ignored).  The
factor of 1/3 is included to highlight that these couplings are
proportional to the baryon number, which is 1/3 for both left- and
right-handed quarks.  Furthermore, we consider a leptophobic $Z'$, so
its tree-level couplings to leptons are also proportional to the
baryon number, which is 0 for leptons.  We use the label $Z'_B$ for
the $Z'$ boson that has the couplings proportional to the baryon
number. 

We emphasize, though, that baryon number does not play any significant role in this Section.
The flavor-independent couplings (\ref{eq:quark-couplings}) are considered here because they are convenient for comparing the many existing hadron collider limits without having to analyze constraints from flavor-changing processes. Furthermore, the collider limits on $Z'_B$ discussed later in this Section depend mostly on the couplings to the $u$ and $d$ quarks, because the parton-distribution functions (PDFs) of the other quarks are much smaller.
Adapting these collider limits on $Z'_B$ to $Z'$ bosons that have different couplings to the $u$ and $d$ quarks is also relatively straightforward,  
by a rescaling of the $u/d$ PDF ratio.

The theory that includes $Z'_B$, which is a massive spin-1 particle,
is well-behaved at high-energies only if $Z'_B$ is a gauge boson or a
bound state. Either way, additional fields must be present.  Here we
will assume that any such fields that couple to $Z'_B$ are
sufficiently heavy (usually above $M_{Z'}/2$), so that the only
tree-level 2-body decays of the $Z'_B$ boson are induced by
Eq.~(\ref{eq:quark-couplings}). 

To be more specific, we will focus on the case where $Z'_B$ is the
gauge boson associated with a $U(1)_B$ symmetry.  Since $Z'_B$ is
massive, there must be a $U(1)_B$ symmetry breaking sector.  The
simplest choice is a complex scalar $\phi$ that is a SM gauge singlet
and carries $U(1)_B$ charge. In addition, there must be some set of
new fermions (``anomalons'') charged under $SU(2)_W \times U(1)_Y
\times U(1)_B$ such that all gauge anomalies cancel~\cite{Dobrescu:2013coa, Duerr:2013dza, Dobrescu:2014fca, Dobrescu:2015asa}.  
Thus, the full Lagrangian of the renormalizable model discussed here comprises,
besides the SM, the following sectors: the kinetic terms for $Z'$
[which includes the interaction terms (\ref{eq:quark-couplings})],
$\phi$, and each anomalon, the potential for $\phi$ that spontaneously
breaks $U(1)_B$, as well as Yukawa interactions of two anomalons with
$\phi$, and of one anomalon and one SM fermion with the SM Higgs
doublet.

We assume that the anomalon masses, which are mostly induced by $\phi$, are heavier than $M_{Z'}/2$. The opposite case, where the anomalon masses are lighter 
than $M_{Z'}/2$, has a highly model-dependent phenomenology due to the cascade decays of $Z'_B$ via pairs of anomalons \cite{Dobrescu:2017sue}.
Besides decays to anomalons, a $Z'$ boson could in principle decay into additional particles beyond the SM, as studied for example in \cite{Araz:2017wbp}; we will not consider that possibility in this work.

There are two types of $Z'_B$ decay modes at tree-level: into two jets, and into $t\bar t$ if $M_{Z'} > 2 m_t$.
The branching fraction into two jets is given at leading order by
\be
B \! \left( Z'_B \to jj \right) =  \left[ 1 + \dfrac{1}{5}  \left( 1 + \dfrac{2 m_t^2 }{M_{Z'}^2 } \right)\left( 1 - \dfrac{4 m_t^2 }{M_{Z'}^2 } \right)^{\! \! 1/2} \,  \right]^{\! -1} ~~.
 \label{eq:Br}
\ee
This branching fraction\footnote{The $m_t/M_{Z'}$ dependence here
  corrects a typo from Eq. (7) of Ref.~\cite{Dobrescu:2013coa}.}
approaches 5/6 for $M_{Z'} \gg 2 m_t$, and 1 for $M_{Z'} \lesssim 2
m_t$.  The ratio between the total width and mass of the $Z'_B$ boson
is $\Gamma_{Z'}/M_{Z'} \approx g_{_B}^2/(24\pi) $ for $M_{Z'} \gg 2
m_t$, and is 5/6 of that for $M_{Z'} \lesssim 2 m_t$.

The properties of $Z'_B$ primarily depend on two parameters: the mass
$M_{Z'}$ and the coupling $g_{_B}$. It is natural, therefore, to
present the collider limits in the $(M_{Z'}, g_{_B})$
plane~\cite{Dobrescu:2013coa}.  The $s$-channel production cross
section of $Z'_B$ at hadron colliders is proportional to $g_{_B}^2$
and quickly decreases with $M_{Z'}$. At leading order, $Z'_B$
production proceeds from quark-antiquark initial states. At
next-to-leading order (NLO) in QCD, there are also contributions from
quark-gluon initial states. We have computed the $Z'_B$ NLO production 
cross section at the LHC using the MadGraph\_aMC@NLO
code~\cite{Alwall:2014hca}, with model files generated by
FeynRules~\cite{Alloul:2013bka} (which uses the FeynArts
package~\cite{Hahn:2000kx} for NLO corrections), and 
the PDF set NNPDF3.1 NLO \cite{NNPDF:2017mvq} with $\alpha_s(M_Z) = 0.118$.
The MadGraph\_aMC@NLO default dynamical factorization and renormalization scale (which is determined by the $p_T$ of the decay products) was used, so that $\alpha_s$ is evaluated at a scale that is event-dependent. 

 \begin{figure}[t!]
  \begin{center}
 \includegraphics[width=0.75\textwidth, angle=0]{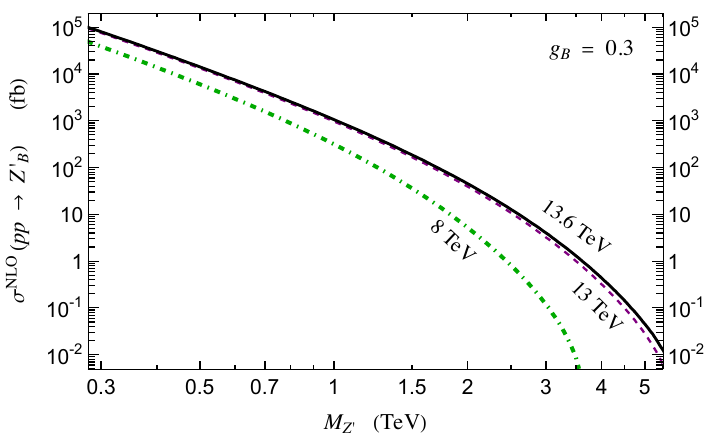}
   \caption{Cross section for production of a $Z'_B$ boson in
     proton-proton collisions at $\sqrt{s} = 13.6$ TeV (solid black line), 13 TeV (dashed purple line), and 8 TeV (green dash-dotted line).  The
     coupling is fixed here at $g_{_B} = 0.3$; the cross section
     scales as $g_{_B}^2$.  Next-to-leading order QCD corrections are
     included through MadGraph\_aMC@NLO.  }
  \label{fig:xsec}
  \end{center}
\end{figure}

The resulting cross sections are shown in Figure~\ref{fig:xsec} for a $Z'$ gauge coupling fixed at $g_{_B} = 0.3$.
This value has been chosen for illustrative purposes; 
note that $g_{_B}$ is a free parameter of order one or smaller, and that the $Z'$ production cross sections scale as $g_{_B}^2$. 
The cross sections shown in Figure~\ref{fig:xsec}  are computed for 
center of mass energies of 13.6 TeV (the current one in Run 3 of the LHC), 13 TeV (used in Run 2) and 8 TeV (used in Run 1). 
The cross section is larger at $\sqrt{s} = 13.6$ TeV than at $\sqrt{s} = 13$ TeV by a factor that grows from 5\% at $M_{Z'} = 0.1$ TeV to 9\% at $M_{Z'} = 1$ TeV and 36\% at $M_{Z'} = 4$ TeV. 

\begin{figure}[t]
  \begin{center}
  \hspace*{-1.5mm} 
 \includegraphics[width=0.495\textwidth, angle=0]{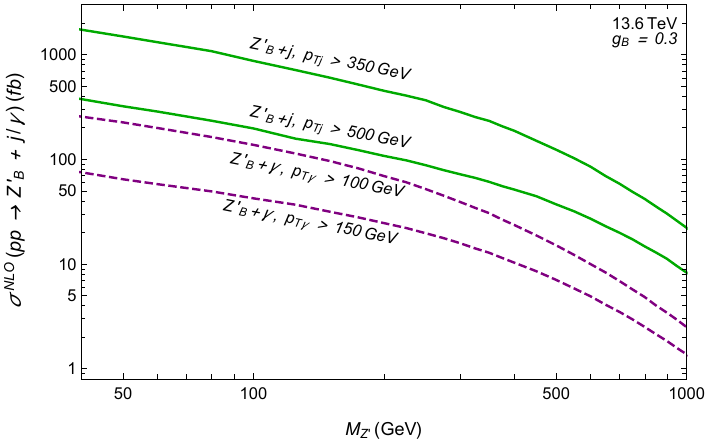}\ 
 \includegraphics[width=0.495\textwidth, angle=0]{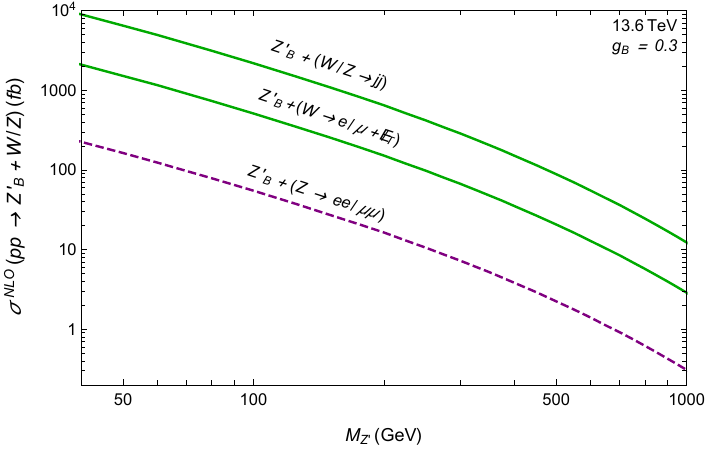}
   \caption{Next-to-leading order cross section for a $Z'_B$ boson
     produced at the 13.6 TeV LHC in association with a jet (solid blue
     lines) or photon (dashed purple lines) of transverse momentum
     above a certain limit (left panel), or with a weak boson decaying
     to jets or leptons (right panel).  The coupling used here is
     $g_{_B} = 0.3$; all cross sections scale as $g_{_B}^2$. }
  \label{fig:xsec2}
  \end{center}
\end{figure}

The most stringent collider limits on the coupling for $M_{Z'} < 450$
GeV are set by LHC searches for a dijet resonance produced in
association with an initial state jet, photon or a
leptonically-decaying $W$ boson. The production cross sections for
$Z'_B j$ and $Z'_B \gamma$ at the 13.6 TeV LHC, computed at NLO with
MadGraph\_aMC@NLO, are shown in the left-hand panel of
Figure~\ref{fig:xsec2} for two choices of the $p_T$ cut on the initial
state radiation (ISR).  The $Z'_B W$ production cross section times
the $W$ branching fraction into leptonic final states (excluding $\tau
\nu$) is given in the right-hand panel of Figure~\ref{fig:xsec2}, by
the dashed gray line.

We point out that additional processes that can be used in future
searches at low dijet mass involve initial state radiation of a $W$ or
$Z$ boson decaying hadronically. The cross section for $Z'_B$
production in association with an electroweak boson that decay into
jets is shown by the solid red line in the right-hand panel of
Figure~\ref{fig:xsec2}.  Note that at low mass this rate is larger by
a factor of about 5 than the $Z'_B j$ rate with $p_{Tj} > 350$ GeV, so
searches for associated $Z'_B \, W/Z$ production appear promising.
The $Z'_B Z$ production cross section times the sum of $Z$ branching
fractions into $e^+e^-$ and $\mu^+ \mu^-$ is also given in the
right-hand panel of Figure~\ref{fig:xsec2} (see the dotted blue line).
The low background for events with a leptonically-decaying $Z$ and a
$jj$ resonance would allow the use of $Z'_B Z$ production to improve
the sensitivity to lower dijet masses.

Using an ISR jet as a trigger for light dijet resonances has been a
key aspect for the current search sensitivity at low
masses~\cite{CMS:2017nuu, CMS:2017dcz, ATLAS:2018hbc}.  As
a practical matter, however, the large boost to the $Z'_B$ resonance
necessitates the use of jet substructure techniques to both remove
contamination from pile-up and distinguish the $Z'_B$ peak signal from
the overwhelming QCD background. The $p_T$ requirement from the ISR
jet~\cite{CMS:2017nuu, CMS:2017dcz, ATLAS:2018hbc} thus
leads to a sculpted invariant mass distribution, necessitating the use
of novel experimental techniques to decorrelate the $p_T$ of the ISR
jet from the differential mass distribution~\cite{Dolen:2016kst}.

 \begin{figure}[t]
  \begin{center}
  \hspace*{-.3cm} 
  \includegraphics[width=0.99\textwidth, angle=0]{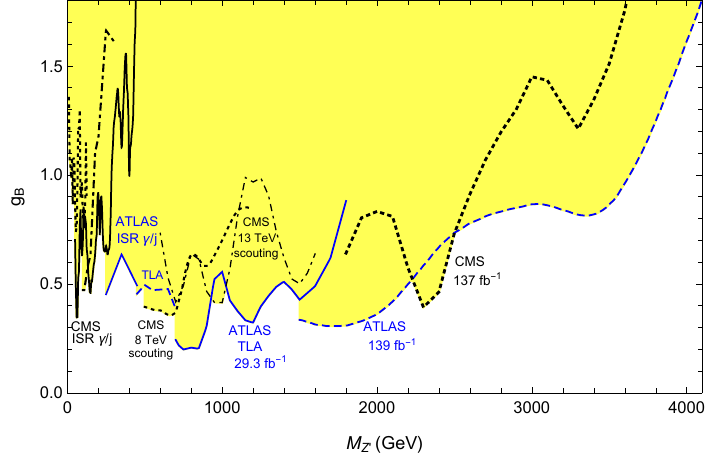}
   \caption{ Limits on the $Z'_B$ boson in the coupling-mass plane,
     based on the ATLAS and CMS searches~\cite{CMS:2016ltu,
       CMS:2017nuu, ATLAS:2018qto, CMS:2018mgb, ATLAS:2019itm,
       CMS:2019xai, CMS:2019emo, ATLAS:2019fgd, CMS:2019gwf} described
     in the text. The yellow shaded region is excluded at the 95\%
     confidence level. The normalization of the $g_{_B}$ coupling used
     here is given in Eq.~(\ref{eq:quark-couplings}).  }
  \label{fig:bounds}
  \end{center}
\end{figure}

The current coupling-mass limits are shown in Figure~\ref{fig:bounds},
and are derived from various types of hadron collider searches,
depending on the resonance mass.  Only searches that set the most
stringent limits for some mass range are included
there~\cite{CMS:2016ltu, CMS:2017nuu, ATLAS:2018qto, CMS:2018mgb,
  ATLAS:2019itm, CMS:2019xai, CMS:2019emo, ATLAS:2019fgd,
  CMS:2019gwf}.  Earlier limits that have been superseded can be found
in~\cite{Dobrescu:2013coa}.

For $M_{Z'} > 1.5$ TeV, the most stringent limits on $g_{_B}$ are set
by dijet resonance searches at $\sqrt{s} =13$~TeV.  The CMS
search~\cite{CMS:2019gwf} is based on the full Run 2 luminosity,
totaling 137 fb$^{-1}$ of data (which supersedes the earlier high-mass
results~\cite{CMS:2018mgb, CMS:2018wxx}).  The ATLAS
search~\cite{ATLAS:2019fgd} is also based on the full Run 2
luminosity, totaling 139 fb$^{-1}$ of data.

These limits assume that the dijet signal is given by the $Z'_B$
production cross section times the branching fraction $B \! \left(
Z'_B \to jj \right) $ given in Eq.~(\ref{eq:Br}). In practice, for
$M_{Z'} \gg 2 m_t$ there is an additional contribution from $Z'_B \to
t \bar t$ because each top quark is highly boosted and may appear as a
jet. This effect is weaker in the case of ATLAS searches, where the
jet cone size is $R = 0.5$, significantly smaller than the one used in
CMS dijet searches, $R = 1.1$.  The $ t \bar t$ invariant mass
distribution matches the dijet one only when both top quarks decay
hadronically, otherwise the neutrinos shift the invariant mass below
$M_{Z'}$.  Thus, the effective dijet branching fraction of $Z'_B$ is
slightly higher than 5/6, reaching $5/6 + B(W\to jj)^2/6 \approx 0.91$
at very high masses.  Consequently, the limits on $g_{_B}$ may be up
to 5\% tighter than those shown in Figure~\ref{fig:bounds}.

For 450~GeV $ < M_{Z'} < 1.5$~TeV, there is competition between four
searches.  The CMS method in that mass range, called ``scouting'',
uses dijets reconstructed from calorimeter information in the trigger.
The latest CMS search of this type uses 27 fb$^{-1}$ of 13 TeV data
(the low-mass result of~\cite{CMS:2018mgb}) and sets a competitive
limit especially in the 0.9--1 TeV mass range, while the search with
18.8 fb$^{-1}$ of 8~TeV data~\cite{CMS:2016ltu} still sets the most
stringent limit in the 500--700 GeV mass range.  The similar ATLAS
``Trigger-Level Analysis'' (TLA) used 29.3 fb$^{-1}$ of 13~TeV
data~\cite{ATLAS:2018qto}, setting the most stringent limit in the
700--900 GeV and 1--1.5 TeV mass ranges; a version of that
search~\cite{ATLAS:2018qto} with a different event selection and only
3.4 fb$^{-1}$ sets the most stringent limit in the 450--500 GeV mass
range.

In the 237--450 GeV mass range, the best limits are set by the ATLAS
searches with initial state radiation (ISR) of a jet or a photon (79.8
fb$^{-1}$ at 13 TeV~\cite{ATLAS:2019itm}) and $b$-tagged jets.  The
ATLAS 139 fb$^{-1}$ search using an ISR $W$ boson giving a high-$p_T$
lepton for the trigger~\cite{Aad:2020kep} gives a slightly weaker bound.

Finally, for 50--237 GeV, the limit is set for most $Z'_B$ masses by
the CMS search for a dijet resonance plus an ISR jet with 35.9
fb$^{-1}$ accumulated in 2016~\cite{CMS:2019emo}, where the dijet
system is boosted and merged into a single jet with substructure.  In
the relatively narrow 100--135 GeV mass range, the strongest limit is
set by the similar CMS search with the 2015 data of 2.7
fb$^{-1}$~\cite{CMS:2017nuu}.  From 10--50 GeV, the CMS analysis with
an ISR photon with 35.9 fb$^{-1}$~\cite{CMS:2019xai} gives the leading
direct constraint on dijet resonances.

The coupling-mass plot of Figure~\ref{fig:bounds} shows that there is
a gap in sensitivity for $M_{Z'}$ roughly in the 200--500 GeV
range. Improved techniques will be required to fill that gap. By
contrast, the high-mass region will continue to be covered by existing
analyses applied to larger data sets.  Higher-energy proton-proton
colliders will substantially increase the reach at high
$M_{Z'}$~\cite{Yu:2013wta}.

At the other end of the plot, masses below 100 GeV are also
constrained by electroweak precision and quarkonium measurements.  We
will next derive these constraints by calculating the mixing of the
$Z'$ with SM states.

\section{$Z'_B$ mixing with the $Z$ and the photon}
\setcounter{equation}{0}
\label{sec:mixing}

Besides the limits from hadron collider searches discussed in the
previous Section, there are constraints on the $Z'_B$ boson from
measurements of the $Z$ boson properties.  The mixing between the SM
$Z$ boson (labelled here $Z_{\rm SM}$) and the $Z'_B$ boson may modify
the branching fractions of the observed $Z$ boson compared to the SM
predictions at a level incompatible with existing measurements.
Furthermore, the $Z'_B - Z_{\rm SM}$ mixing as well as the kinetic
mixing between the $Z'_B$ and the photon lead to $Z'_B$ decays into
pairs of leptons~\cite{Dobrescu:2017sue}, which are constrained by
searches for dilepton resonances.

A kinetic mixing between the SM hypercharge gauge boson and the $Z'_B$
may in principle be present at tree level. If, however, the $U(1)_Y$
or the $U(1)_B$ gauge groups are embedded in a non-Abelian structure
at some high scale (which is generically expected as they are not
asymptotically free), then the tree-level kinetic mixing vanishes.

Nevertheless, a $Z'_B-Z_{\rm SM}$ mixing will be generated by loops
involving fields that couple to both bosons.  To compute the 1-loop
$Z'_B-Z_{\rm SM}$ mixing, we first need to specify all the fields
that carry electroweak charges and also couple to $Z'_B$.

\subsection{Kinetic mixing}

Let us consider in what follows the theory where $Z'_B$ is the gauge
boson associated with a $U(1)_B$ gauge symmetry, so that $Z'_B$ does
not couple to leptons while its couplings given in
Eq.~(\ref{eq:quark-couplings}) arise when all SM quarks carry the same
$U(1)_B$ charge, chosen to be 1/3.  The gauge theory with these quark
charges is not self-consistent unless certain new fermions, called
anomalons, are present to cancel the gauge anomalies. If some of these
have masses below $M_{Z'}/2$, then the $Z'_B$ can decay into anomalon
pairs, leading to interesting collider
signatures~\cite{Dobrescu:2015asa}.

We focus on sets of anomalons which together with the SM quarks
satisfy the orthogonality condition Tr$(Y B) = 0$, where the trace is
over all fields, $Y$ is the hypercharge, and $B$ is the $U(1)_B$
charge. More explicitly, the condition is
\be
\sum_{f = \rm quarks, \; anom.}  \!  N_f \left( Y_L^f B_L^f + Y_R^f
B_R^f \right) = 0 ~~,
\label{eq:trace}
\ee
where the sum is over the fermions $f$, which are all the anomalons
and the SM quarks in the gauge eigenstate basis, $Y_L^f$ and $B_L^f$
are the hypercharge and $U(1)_B$ charge, respectively, of the
left-handed $f$ fermion, and $Y_R^f$ and $B_R^f$ are the corresponding
charges of the right-handed fermions.  The color factor is $N_f = 3$
when $f$ is a quark, and $N_f = 1$ when $f$ is a color-singlet
fermion.  When the above equation is satisfied, the leading 1-loop
contribution to the kinetic mixing between the SM hypercharge gauge
boson and the $Z'_B$ vanishes, so the constraint from $Z$ measurements
is weak.

Kinetic mixing operators are still generated at one loop due to the
mass differences between anomalons and SM quarks.  The leading
operators of this type~\cite{Dobrescu:2017sue} have dimension six and
involve the SM Higgs doublet, $H$, or the $U(1)_B$-breaking scalar
$\phi$:
\be
\phi^\dagger \phi \,  Z'_{B\mu \nu} B^{\mu \nu} \;\; ,  \;\;  H^\dagger H \,  Z'_{B\mu \nu} B^{\mu \nu}  \;\;  ,  \;\; H^\dagger  \tau^a 
 H \,  Z'_{B \mu \nu} W^{a \mu \nu}  ~~, 
\ee 
where $B^{\mu \nu}$ and $W^{a \mu \nu}$ are the hypercharge and
$SU(2)_W$ field strengths.

There are also mass mixing operators, which arise at dimension six: 
\be
H^\dagger (D^\mu H) \, \phi  \, (D_\mu \phi^\dagger) + {\rm H.c.}   \;\;  ,   \;\; 
H^\dagger (D^\mu H) \phi^\dagger  (D_\mu \phi) + {\rm H.c.}  \ .
\label{eq:massOps}
\ee 
These may arise at one loop, depending on the anomalon charges. Once a
Higgs doublet and a $ \phi $ scalar are replaced by their vacuum expectation values (VEVs), a
$Z'_B - Z_{\rm SM}$ mass mixing is induced.

As the masses of the $Z'_B$ and the anomalons may be at or below the
weak scale, it is appropriate to compute the mixings of $Z'_B$ with
the $Z_{\rm SM}$ and the photon rather than the ones involving the
$SU(2)_W \times U(1)_Y$ gauge bosons. The Lagrangian terms for these
can be written as
\be
 \frac{ 1}{2} \, Z_B^{\prime \, \mu \nu} \left(  \km_Z \, Z_{\rm SM \mu \nu} -  \km_\gamma  F_{\mu \nu}  \right) 
 +  \Delta M^2_{Z'Z}  Z_B^{\prime \, \mu} Z_{\rm SM \mu}   ~~,
\label{eq:LagrMix}
\ee 
where the coefficients $\km_Z$ and $\km_\gamma$ are dimensionless and
real, and $\Delta M^2_{Z' Z} $ is a mass squared parameter.  The field
strengths for $Z'_B$, $Z_{\rm SM}$ and the photon are canonically
normalized, {\it i.e.}, the tree-level kinetic terms are $(-1/4)
(Z_B^{\prime \, \mu \nu} Z'_{B\mu \nu} + Z^{\mu \nu}_{\rm SM} Z_{\rm
  SM \mu \nu} + F^{\mu \nu} F_{\mu \nu}) $.

The real part of the $Z'_B-Z_{\rm SM}$ mixing amplitude contains two
pieces: a kinetic mixing and a mass mixing.  The $Z'_B-Z_{\rm SM}$
mixing amplitude can be written as $\epsilon_\mu (Z'_B) \epsilon_\nu
(Z) {\cal A}^{\mu\nu}_{Z'Z} $, with $\epsilon_\mu (Z'_B) $ and $
\epsilon_\nu (Z)$ being the polarization vectors of the two gauge
bosons. The real part of ${\cal A}^{\mu\nu}_{Z'Z} $ is
\be
{\rm Re} \, {\cal A}^{\mu\nu}_{Z'Z} = \km_Z \left( g^{\mu\nu} p^2 -
p^\mu p^\nu \right) + \Delta M^2_{Z'Z} \, g^{\mu\nu} ~~,
\label{eq:Amunuform}
\ee 
where $p^\mu$ is the 4-momentum of the $Z'_B$ or $Z_{\rm SM}$ bosons.

The sine and cosine of the weak mixing angle are labelled in what
follows by $s_W$ and $c_W$, while the $SU(2)_W$ gauge coupling is $g =
e/s_W$, where $e$ is the electromagnetic gauge coupling.  Expressing
the $Z_{\rm SM}$ couplings of the left- and right-handed fermion $f$
(without the $g/c_W$ prefactor) in terms of their $T^3$ value and
hypercharge,
\be
g_{L,R}^f = c_W^2 T^3_{L,R} - s_W^2 Y^f_{L,R}  ~~,
\label{eq:gLgR}
\ee
we find that the sum of the $Z_{\rm SM}$ couplings over the fermions
belonging to an $SU(2)_W$ multiplet of size $n$ is proportional to the
hypercharge $Y^f$ of that representation:
\be
\sum_{f \in \, \rm n} g_{L,R}^f = - n \, s_W^2 Y^f_{L,R}   ~~.
\label{eq:gsumY}
\ee
From Eq.~(\ref{eq:trace}) then follows an important sum rule:
\be
\sum_{f = \rm quarks, \; anom.}  N_f \left( g_L^f B_L^f + g_R^f B_R^f
\right) = 0 ~~.
\label{eq:sumRule1}
\ee

If all the SM quarks and anomalons had the same mass, then
Eq.~(\ref{eq:sumRule1}) would have implied that the $Z'_B-Z_{\rm SM}$
kinetic mixing vanishes at one loop.  As the top quark is much heavier
than the other SM quarks, the kinetic mixing receives a significant
contribution from the SM.  The anomalons also contribute to the
kinetic mixing, with an amount sensitive to the anomalon masses and
also to the anomalon charges.  The dependence of the kinetic mixing on
the anomalon set has not been recognized in previous
work~\cite{Carone:1994aa, Carone:1995pu, Aranda:1998fr,
  Graesser:2011vj, Dobrescu:2014fca}.  Similarly, the fact that the
loop-induced kinetic mixing is finite has been mostly overlooked (an
exception is~\cite{Dobrescu:2017sue}).

To be concrete, we analyze a renormalizable Lagrangian that includes
the SM plus the canonical kinetic terms for the $U(1)_B$ gauge boson,
for a complex scalar $\phi$ of $U(1)_B$ charge $+3$, and for a minimal
set of anomalons that satisfies the trace
condition~(\ref{eq:sumRule1}), as well as a $\phi$ potential and
Yukawa couplings.  There are no tree-level kinetic or mass mixings
involving the fields beyond the SM.  Fermion loops will generate
kinetic mixing, but no independent $\Delta M^2_{Z' Z}$ mass mixing
because the anomalons are vectorlike with respect to the SM gauge
group.

In the Appendix, we compute the mixing between $Z_{\rm SM}$ and any
$Z'$ induced at one loop by any fermions that satisfy an orthogonality
relation like~(\ref{eq:sumRule1}).  In this section we are primarily
interested in the case where the 4-momentum of the gauge bosons
satisfies $p^2 = M_Z^2$, so that we can extract limits on the $Z'$
from measurements at the $Z$ pole.  The 1-loop computations of the
mixings are simplified when the anomalon couplings to the Higgs
doublet are negligible, {\it i.e.}, the anomalon masses come entirely
from Yukawa couplings to the scalar $\phi$ responsible for
spontaneously breaking $U(1)_B$.  In that situation there are no
1-loop contributions to $\Delta M^2_{Z'Z} $ because the operators
(\ref{eq:massOps}) cannot be generated either by SM quarks (which do
not couple to $\phi$) or by anomalons (which do not couple to $H$).
This can also be seen from (\ref{eq:kappa}), which gives $\Delta
M^2_{Z'Z} $ after setting $z^q_L = z^q_R$ for the SM quarks and $g_L^f
= g_R^f $ for the anomalons.

The expansion in (\ref{eq:Fylow}) shows that the loops involving the
SM quarks other than the top quark have contributions to the kinetic
mixing which are of order $(m_q/M_Z)^2$ where $m_q$ are the SM quark
masses, and thus can be neglected.  Hence, the kinetic mixing, given
in general in (\ref{eq:kappaZ}), becomes a sum over the top quark and
anomalon contributions:
\be 
\hspace*{-0.2cm}
 \km_Z   \simeq   \frac{ g_{_B} g}{ 48 \pi^2 c_W } \,  \left[    \left( \frac{1}{2} - \frac{4}{3} s_W^2 \right)   {\cal F}(m_t^2/M_Z^2)  + 
\!\!  \sum_{f=\rm anom.} \!\!  N_f  \left( g_L^f  B_L^f + g_R^f  B_R^f  \right)  
  {\cal F}(m_f^2/M_Z^2)  \right]  ,
  \label{eq:kinMix}
\ee
The function ${\cal F}$ is  given in Eq.~(\ref{eq:Fyfull}) of the Appendix,
and for  $m_f \gtrsim M_Z$ is well approximated by 
\be
 {\cal F}(m_f^2/M_Z^2)  \simeq   2 \, \ln \! \left(\frac{m_f}{M_Z} \right) + \frac{5}{3} - \frac{M_Z^2}{5 \, m_f^2}  ~~.
\ee
For $m_f$ in the interval 100--400 GeV, ${\cal F}(m_f^2/M_Z^2)$
continuously grows from 1.67 to 4.61.

As mentioned in Section~\ref{sec:coupling-mass}, we will focus here on
the case where all the anomalons are color singlets ($N_f = 1$) and
heavier than $M_{Z'}/2$, where $M_{Z'}$ is the mass of the physical
particle $Z'$.  The collider constraints on the anomalons are weak in
this case: pair production at LEP II sets a lower limit on the
anomalon mass of about 90 GeV, depending on the anomalon decay
modes~\cite{Dobrescu:2014fca}.  Using (\ref{eq:gsumY}) and replacing
the known quantities in Eq.~(\ref{eq:kinMix}) by their numerical
values, we find the following expression for the $Z'_B-Z_{\rm SM}$
kinetic mixing at one loop:
\be
\km_Z  \simeq   8.70 \times 10^{-4}  \, g_{_B}  \left( 1 - 0.417  \sum_{f=\rm anom.} \!  Y^f  \! \left(  B_L^f +  B_R^f  \right)  
  {\cal F}(m_f^2/M_Z^2) \right)    ~~.
\label{eq:kappaZap}
\ee
The same computation detailed in the Appendix, but with
the $Z$ couplings replaced by the photon ones, gives the following expression for the kinetic mixing of the $Z'_B$ with the SM photon, defined in (\ref{eq:LagrMix}):
\be  \hspace*{-0.2cm}
 \kappa_\gamma   \simeq   \frac{ -g_{_B}  e}{ 48 \pi^2 } \, \left[ \frac{4}{3}   {\cal F}(m_t^2/M_{Z'}^2)  + 
\!\!  \sum_{f=\rm anom.} \!\!     Q^f  \left(  B_L^f +  B_R^f  \right)  
  {\cal F}(m_f^2/M_{Z'}^2)  \right]  .
  \label{eq:kinMixgamma}
\ee 

A minimal set of anomalons which includes only color singlets, cancels
all gauge anomalies, and satisfies the trace condition is given by the
following $SU(2)_W \times U(1)_Y \times U(1)_B$
representations~\cite{Duerr:2013dza, Dobrescu:2014fca, Dobrescu:2015asa}
\bear
&& L_L (2,-1/2, -1) \;\; , \;\; L_R (2, -1/2, 2) \;\; , \;\; E_L (1, -1, 2) \;\; , \;\; E_R (1, -1, -1)
\nonumber \\ [2mm] 
&& N_L (1, 0, 2) \;\; , \;\; N_R (1, 0, -1) \ .
\label{eq:anomSet}
\eear
The SM gauge singlet fermions, $N_L$ and $N_R$, are required to cancel
the $U(1)_B$ and $[U(1)_B]^3$ anomalies, but do not contribute to the
kinetic mixing.  The anomalons acquire mass from the scalar $\phi$,
with $U(1)_B$ charge $+3$ and whose VEV
$\langle \phi \rangle = v_\phi$ breaks the $U(1)_B$ symmetry. The
corresponding Yukawa interactions
\be 
- y_L \overline{L}_L \phi^* L_R - y_E \overline{E}_L \phi E_R - y_N
\overline{N}_L \phi N_R + \text{ H.c.} \ ,
\label{eqn:anomalons}
\ee 
set the anomalon masses to be $y_L v_\phi $, $y_E v_\phi$, and $y_N
v_\phi$.  We assume the dominant mass generation arises from $U(1)_B$
breaking, and neglect the possible Yukawa interactions to the SM Higgs
doublet.  We remark that small Yukawa interactions to the SM Higgs
doublet, which are needed to ensure the charged anomalons can decay to
SM fermions, are still allowed by $h \to \gamma \gamma$
constraints~\cite{Michaels:2020fzj}.  
These and other Higgs observables also exclude some of the original
models for local baryon number~\cite{Carone:1994aa, Bailey:1994qv}.
If all anomalons have the same mass, $m_f \gtrsim 90$ GeV, then
the anomalon-dependent factor in (\ref{eq:kappaZap}) becomes
\be
  \sum_{f=\rm anom.} \!  Y^f  \left( B_L^f +   B_R^f  \right)   {\cal F}(m_f^2/M_Z^2)  = - 2  \, {\cal F}(m_f^2/M_Z^2)   ~~,
\ee
and we obtain that this anomalon set gives $\km_Z/g_{_B} \simeq 2.08
\times 10^{-3} $ for $m_f = 100$ GeV, and $\km_Z/g_{_B} \simeq 3.19
\times 10^{-3} $ for $m_f = 200$ GeV.  Under the same assumptions, the
photon kinetic mixing in (\ref{eq:kinMixgamma}) with $M_{Z'} \approx
M_Z$ gives $\kappa_\gamma/g_{_B} \simeq -3.32 \times 10^{-4}$ for $m_f
= 100$~GeV, and $\kappa_\gamma/g_{_B} \simeq 1.69 \times 10^{-3}$ for
$m_f = 200$~GeV. The values for $\kappa_\gamma$ and $\kappa_Z$ are
roughly comparable because both originate from a single kinetic mixing
of the hypercharge gauge field with the $Z’_B$ field.


\subsection{Couplings of the physical bosons}

We now diagonalize the kinetic terms for the $Z_{\rm SM}$, $Z'_B$
bosons and the photon, including the mixing terms from
(\ref{eq:LagrMix}).  Given that the kinetic mixing with the photon has
only a subdominant impact on phenomenology (due to the tree-level
couplings of $Z'_B$ to quarks), it is convenient to work in the
leading order in $\kappa_\gamma \ll 1$. It is then sufficient to
redefine $Z_{\rm SM}$ and $Z'_B$ first to absorb the kinetic mixing
$\kappa_Z$, where the non-unitary nature of the field redefinition
induces mass mixing between the two heavy bosons.  The induced mass
mixing is symmetric and requires one rotation angle to obtain diagonal
mass eigenstates.  The kinetic mixing with the photon is absorbed by a
redefinition of the photon field by $\kappa_\gamma Z'_B$, which leads
to $Z'_B$ couplings to the electromagnetic current proportional to
$\kappa_\gamma$ and no further mass mixing, as studied in
Ref.~\cite{Dobrescu:2017sue}.  A more general diagonalization of the
kinetic mixing between $Z_{\rm SM}$, $Z'_B$, and the photon can be
found in Ref.~\cite{Liu:2017lpo}.

Combining the field redefinition of $Z_{\rm SM}$ and $Z'_B$ and mass
diagonalization attributed to $\kappa_Z$, we find that the mass
eigenstate bosons, labelled by $Z$ and $Z'$, are
\bear
Z^\mu = \cos \theta \, Z^\mu_{\rm SM} +    \left(\sin \theta  \,  \sqrt{1 -  \kappa_Z^2 } -  \kappa_Z  \cos \theta  \right)  Z^{\prime \mu}_B     ~~,
\nonumber \\ [2mm]
Z^{\prime \mu}   =   \left(\cos \theta   \, \sqrt{1 -  \kappa_Z^2 }+ \kappa_Z  \sin \theta  \right)  Z^{\prime \mu}_B   - \sin \theta \, Z^\mu_{\rm SM}         ~~,
\label{eq:massBasis}
\eear
where  $-\pi/4 < \theta <  \pi/4$ and 
\be
\tan 2\theta = \frac{2  \kappa_Z }{1 -  2 \kappa_Z^2 - M_{Z'_B}^2/M_{Z_{\rm SM}}^2} \sqrt{1 -  \kappa_Z^2 }  ~~.
\label{eq:theta}
\ee
The squared masses of the two physical states are 
\be 
M_{Z,Z'}^2 = \frac{1}{2 (1 - \kappa_Z^2) } \left(  M_{Z_{\rm SM}}^2 + M_{Z'_B}^2 \pm \sqrt{  \left( M_{Z_{\rm SM}}^2 - M_{Z'_B}^2  \right)^2 + 4 \kappa_Z^2 \,  M_{Z_{\rm SM}}^2  M_{Z'_B}^2  } \; \right)  ~~,
\label{eq:squaredMasses}
\ee
where the $+$ sign corresponds to $M_{Z}^2$ only when $M_{Z_{\rm SM}}
\ge M_{Z'_B}$.  Since $\kappa_Z \ll 1$, in what follows we drop the
terms of order $\kappa_Z^2$ from Eq.~(\ref{eq:massBasis}) and from the
prefactor of Eq.~(\ref{eq:squaredMasses}).  As the $Z'_B$ mass may be
close to $M_{Z_{\rm SM}}$, we do not yet expand the denominator of
Eq.~(\ref{eq:theta}) or the last term of Eq.~(\ref{eq:squaredMasses}).

As a consequence of mixing, the couplings of the physical $Z$ boson to
quarks and leptons are changed compared to the SM ones, given in
Eq.~(\ref{eq:gLgR}), as follows
\bear
&& g_{L,R}^q \to  \hat g_{L,R}^q    =     \left(  \cos\theta + \kappa_Z \sin\theta \right) \, g_{L,R}^q + \sin\theta\;  \frac{g_{_B} c_W }{6  \, g}   ~~,
\nonumber \\ [2mm]
&& g_{L,R}^\ell \to  \hat g_{L,R}^\ell    =     \left(  \cos\theta + \kappa_Z \sin\theta \right) \, g_{L,R}^\ell   ~~.
\label{eq:Zcouplings}
\eear
The couplings of the physical $Z'$ boson to quarks are modified
compared to those of $Z'_B$ gauge boson shown in
Eq.~(\ref{eq:quark-couplings}), by a charge- and chirality-dependent
factor:
\be
\cos \theta + (- \sin \theta +  \kappa_Z \cos \theta ) \frac{6g}{g_{_B} c_W} g_{L,R}^q  ~~.
\label{eq:modified-couplings}
\ee
In addition, the $Z'_B-Z_{\rm SM}$ kinetic mixing induces couplings of
$Z'$ to leptons:
\be
Z'_\mu \; \frac{g}{ c_W}  (- \sin \theta +  \kappa_Z \cos \theta )  \sum_\ell \,  \left( g_L^\ell  \, \bar \ell_L  \gamma^\mu \ell_L +  g_R^\ell \,  \bar \ell_R  \gamma^\mu \ell_R \right) ~~.
\ee

The kinetic mixing between $Z'_B$ and the photon, $\kappa_\gamma$,
which is given in (\ref{eq:kinMixgamma}), also contributes to the $Z'$
couplings to leptons, as studied in Ref.~\cite{Dobrescu:2017sue}.
Note, however, that the couplings of $Z'$ to leptons are both
loop-suppressed and proportional to $g_{_B}$, so that the branching
fractions of the $Z'$ into leptons are at the sub-percent level, and
would become relevant only after the $Z'$ discovery via the
quark-antiquark modes.

\subsection{Limits from electroweak measurements}

Let us focus first on the typical case, where the relative mass
splitting of the two gauge bosons is large compared to the kinetic
mixing: $|M_{Z'_B} - M_{Z_{\rm SM}} | \gg \kappa_Z M_{Z_{\rm SM}}$.
In that case Eq.~(\ref{eq:theta}) implies $\sin\theta \ll 1$ and, to
leading order in $\kappa_Z^2$, 
\be
\sin\theta \simeq \frac{ \kappa_Z }{1 -  M_{Z'}^2/M_Z^2}  ~~.
\label{eq:sintheta}
\ee 
Furthermore, the mass difference between the two physical
particles in this case is approximately equal to the mass difference
of the two gauge bosons: $M_{Z'} - M_Z \simeq M_{Z'_B} - M_{Z_{\rm
    SM}} $ up to corrections of order $(\kappa_Z M_{Z_{\rm SM}})^2/
(M_{Z'_B} - M_{Z_{\rm SM}} )^2$.  The constraints from $Z$ pole
measurements depend on the size of the $M_{Z'} - M_Z$ mass splitting
compared to the measured $Z$ width, $\Gamma_Z \approx 2.5$ GeV.

When $|M_{Z'} - M_Z| \gtrsim \Gamma_Z$, the contribution from $Z'$
exchange to the $Z$ pole observables can be neglected. In that case,
the main effect of the $Z'_B-Z_{\rm SM}$ kinetic mixing is a relative
change in the hadronic $Z$ width compared to the SM prediction:
\bear
\hspace*{-1.2cm}
\frac{\Delta \Gamma_{\rm had}  (Z)}{\Gamma_{\rm had}^{\rm SM}  (Z) }  &\!=\!&    \frac{ 3 \left[ (  \hat  g_{L}^d)^2 + (  \hat  g_{R}^d)^2 \right] + 2  \left[ (  \hat  g_{L}^u)^2 + (  \hat  g_{R}^u)^2 \right] }
{  3 \left[ ( g_{L}^d)^2 + (  g_{R}^d)^2 \right] + 2  \left[ (  g_{L}^u)^2 + (  g_{R}^u)^2 \right]  } - 1  
\nonumber \\ [2mm]
\hspace*{-1.2cm}
 &\! \simeq \!&  - \frac{ A_1 \, g_{_B}  \kappa_Z }{1 -  M_{Z'}^2/M_Z^2} 
  ~~,
\label{eq:hadronic}
\eear
where the coefficient $A_1$ is a function of the weak mixing angle:
\bear
&& A_1 =   \left( \frac{c_W}{6 g} \right)  \frac{1+ 4 s_W^2 /3 }
{  5/4 -  7 s_W^2 /3+ 22 s_W^4/9  }   \; \approx 0.349  ~~~.
\eear
Note that the correction to the leptonic $Z$ width is of order
$\sin^2\!\theta$ and can be neglected here.  For the anomalon set
(\ref{eq:anomSet}), with a common mass fixed at $m_f = 100$ GeV, the
constraint becomes
\be
\frac{\Delta \Gamma_{\rm had}  (Z)}{\Gamma_{\rm had}^{\rm SM}  (Z) }    \simeq  - 7.25 \times 10^{-4} \frac{ g_{_B}^2 }{1 -  M_{Z'}^2/M_Z^2}  ~~.
\label{eq:Delta1}
\ee 
The value for the hadronic $Z$ width obtained from a
fit~\cite{ParticleDataGroup:2022pth} to the LEP I and SLC data is $\Gamma_{\rm
  had} (Z) = 1.7444\pm 0.0020$ GeV, while the SM prediction is
$\Gamma_{\rm had}^{\rm SM} (Z) = 1.7411\pm 0.0008$ GeV. The allowed
interval for the relative change in the hadronic $Z$ width, at the
95\% CL, is
\be
-5.30 \times 10^{-4} <  \frac{\Delta \Gamma_{\rm had}  (Z)}{\Gamma_{\rm had}^{\rm SM}  (Z) }  < 4.30 \times 10^{-3}   ~~.
\label{eq:Zhadronic}
\ee 
Comparing this interval with Eq.~(\ref{eq:Delta1}) leads to the
following upper limit on the $U(1)_B$ gauge coupling:
\be
g_{_B} <  \left\{ \ba{c}  \displaystyle 0.855 \left( 1 -  \frac{M_{Z'}^2}{M_Z^2}  \right)^{\! 1/2}  \; ,  \; {\rm for } \;  M_{Z'}  \lesssim M_Z - \Gamma_Z  ~~, \\ [4mm]
 \displaystyle  2.44  \left(   \frac{M_{Z'}^2}{M_Z^2  } - 1 \right)^{\! 1/2}  \; ,   \; {\rm for } \;  M_{Z'} \gtrsim  M_Z +  \Gamma_Z  ~~.   \ea   \right.  
 \label{eq:limitGeneric}
\ee
assuming the anomalon set (\ref{eq:anomSet}) with a common anomalon mass $m_f = 100$ GeV.
For $m_f = 200$ GeV,  the limit on $g_{_B}$  is multiplied by 0.808.   
For other anomalon charges or masses,
the right-hand side of (\ref{eq:limitGeneric}) is multiplied by 
$(2.08 \times 10^{-3} g_{_B} /\kappa_Z)^{1/2}$, 
where $\kappa_Z$ is given in Eq.~(\ref{eq:kappaZap}).   

When the $Z'$ mass is approximately within one $Z$ width from the $Z$
mass, {\it i.e.,} in the interval 88.7 GeV $ \lesssim M_{Z'} \lesssim
93.7$ GeV, $Z'$ exchange also contributes to processes such as
$e^+e^-\to$ hadrons near the $Z$ pole. In that case the interference
between the $Z$ and $Z'$ exchange amplitudes leads to corrections of
the cross section for $e^+e^-\to$ hadrons near the $Z$ pole,
$\sigma_{\rm had}$, which are not limited to just $\Gamma_{\rm had}
(Z)$. The relative change of $\sigma_{\rm had}$ compared to the SM
prediction is approximately given by
\bear
\hspace*{-1.2cm}
\frac{\Delta \sigma_{\rm had}}{\sigma_{\rm had}^{\rm SM}  }  &\! \simeq  \!&  
  - \frac{ A_1 \, g_{_B}  \kappa_Z }{1 -  M_{Z'}^2/M_Z^2}    \left( 1 - \frac{ \Gamma_Z  \Gamma_{Z'} }{ 4 ( M_Z - M_{Z'})^2 +  \Gamma_{Z'}^2  }  \right)  ~~.
\label{eq:hadronicBoth}
\eear
To derive this we took the energy of the $e^+e^-$ collision to be
$\sqrt{s} = M_Z$. The last term in the parentheses is due to
interference, and depends on the total width of the $Z'$ boson:
$\Gamma_{Z'} \simeq (5/6) g_{_B}^2 M_{Z'} /(24\pi)$ to leading order in
$\sin \theta$.  The fit to the LEP I and SLD data gives $\sigma_{\rm
  had}= 41.541 \pm 0.037$ nb, which is 1.6$\sigma$ higher than the SM
prediction, $\sigma_{\rm had}^{\rm SM} = 41.481 \pm 0.008$
nb~\cite{ParticleDataGroup:2022pth}. As a consequence, the lower limit on
$\sigma_{\rm had}$ is particularly tight at the 95\% CL:
\be
- 3.42 \times 10^{-4} <  \frac{\Delta \sigma_{\rm had} }{\sigma_{\rm had}^{\rm SM} }  < 3.24 \times 10^{-3} \ .
\label{eq:Xsechadronic}
\ee
Comparing this interval with Eq.~(\ref{eq:hadronicBoth}) gives a
nonlinear constraint on $g_{_B}$ as a function of $M_{Z'}$, which
applies to the $M_Z - \Gamma_Z \lesssim M_{Z'} \lesssim M_Z +
\Gamma_Z$ range except for a very narrow region centered around $M_Z$:
\bear
&& \hspace*{-1cm}
g_{_B}^2 -  \! \left[  \left(  \frac{1 - M_{Z'}/M_Z}{ 8.70 \times 10^{-3} \, g_{_B}^2 } \right)^{\! 2} +  0.404 \right]^{-1} \!  < \,  \left\{ 
\ba{c}  \displaystyle 0.944  \left( 1 -  \frac{M_{Z'}}{M_Z}  \right) \, ,  \; {\rm for } \; 
  \kappa_Z  \lesssim 1 - \frac{M_{Z'}}{M_Z }  \lesssim  \frac{\Gamma_Z}{M_Z}   
   \\ [5mm]
 \displaystyle  8.93  \left(   \frac{M_{Z'}}{M_Z  } - 1 \right)  \; ,   \; {\rm for } \;   \kappa_Z   \lesssim  \frac{M_{Z'}}{M_Z} -1   \lesssim  \frac{\Gamma_Z}{M_Z}    
  \ea   \right.  
 \nonumber  \\ [2mm]
&&    \label{eq:nearMZ} 
\eear
Here we used the anomalon set (\ref{eq:anomSet}) with a common mass
fixed at $m_f = 100$ GeV.  We will use the above constraint as well as
Eq.~(\ref{eq:limitGeneric}) when we extend the coupling-mass plot at
low masses in Section 4.  For $m_f = 200$ GeV, the right-hand side of
(\ref{eq:nearMZ}) must be multiplied by a factor of 0.652, while for
other anomalon masses or charges the factor is $2.08 \times 10^{-3}
g_{_B} /\kappa_Z$.

For $|M_{Z'} - M_Z| \lesssim \kappa_Z \, M_Z$, the 1-loop mixing
between $Z'_B$ and $Z_{\rm SM}$ in Eq. (\ref{eq:theta}) is large,
$\sin \theta \approx 1/\sqrt{2}$, as also discussed in
Ref.~\cite{Liu:2017lpo}.  Because the diagonalization to the mass
basis considers only the pole terms in the 1-loop wavefunction
correction, the evaluation of the 1-loop diagrams cannot be neglected
in scattering cross sections.  The mass shift of the $Z_{\rm SM}$
boson from a $Z'$ close in mass was used before as the constraint on
$\kappa_Z$~\cite{Hook:2010tw}, but 
that result needs to be revisited for the very narrow 
region where the relative mass difference is below $ \kappa_Z$.
In particular, the 1-loop interference in
scattering processes with $|M_{Z'} - M_Z| \lesssim \kappa_Z \, M_Z$
leads to interesting new phenomenology akin to neutral meson mixing,
which we reserve for future study.

\section{Low-mass constraints in the minimal $Z'_B$ model}
\setcounter{equation}{0}
\label{sec:low-mass}

The coupling-mass plot is very useful for displaying the LHC dijet
resonance limits. Its linear-linear version (see
Figure~\ref{fig:bounds}), however, does not clearly show the limits
for new bosons at or below the electroweak scale.

\begin{figure}[t!]
  \begin{center}
  \hspace*{-0.2cm} 
  \includegraphics[width=0.99\textwidth, angle=0]{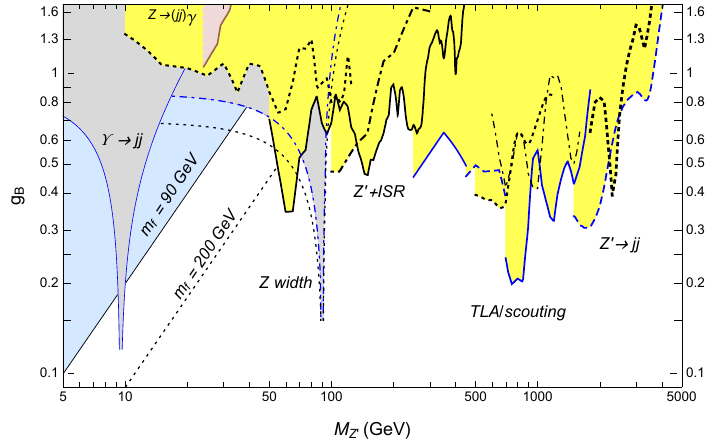}
   \caption{ Limits on the $Z'_B$ boson in the log-log coupling-mass
     plane.  The yellow-shaded region is excluded at the 95\%
     confidence level by dijet resonance searches at the LHC (see
     Figure~\ref{fig:bounds}).  The gray-shaded regions (which are
     particularly strong near 91.2 GeV and 9.5 GeV) are excluded by
     measurements of the $Z$ hadronic width at LEP I (the upper
     dot-dashed line corresponds to an anomalon mass $m_f = 100$~GeV
     and the lower dotted line to $m_f = 200$~GeV), and by
     the ARGUS search for non-electromagnetic $\Upsilon$ decay into a
     jet pair.  The blue-shaded region (above the solid straight line)
     is ruled out by the lower limit on anomalon masses in conjunction
     with the theoretical upper limit on Yukawa couplings, and the
     dotted straight line indicates the possible exclusion if the
     anomalon mass constraint is increased to 200~GeV. The pink-shaded
     region labelled $Z \to (jj) \gamma$ is excluded by the L3 search
     for $Z \to Z' \gamma \to (jj) \gamma$.  }
     \label{fig:bounds2}
  \end{center}
\end{figure}

By contrast, the log-log version of the coupling-mass plot, shown in
Figure~\ref{fig:bounds2}, clearly displays the low mass region.  The
yellow-shaded region is excluded at the 95\% confidence level by dijet
resonance searches at the LHC (and is identical to the shaded region
from Figure~\ref{fig:bounds}).  The gray-shaded region labelled ``$Z$
width'' is ruled out by measurements of the hadronic $Z$ width, which
would be modified by the $Z$--$Z'_B$ mixing induced at one loop by the
SM quarks and also by the anomalons. The boundary of that region is
the limit on the $Z'_B$ coupling, $g_{_B}$, given in
(\ref{eq:limitGeneric}) for a common anomalon mass of 100 GeV.  For
other anomalon masses, the limit changes as described after
(\ref{eq:limitGeneric}), and the bound for a common anomalon mass of
200 GeV is shown as a dotted line for concreteness.  The limit is more
complicated [see (\ref{eq:nearMZ}) and the text after that] when
$|M_{Z'} - M_Z| \lesssim \Gamma_Z$, due to interference effects in the
cross section for $e^+e^-\to$ hadrons.

The gray-shaded region labelled ``$\Upsilon \to jj $'' in
Figure~\ref{fig:bounds2} is excluded by the search for
non-electromagnetic decays of $\Upsilon$ into a jet pair performed by
the ARGUS Collaboration~\cite{Albrecht:1986ec}.  This constraint is
related to the ratio $R_\Upsilon = \Gamma(\Upsilon \to \text{
  hadrons})/\Gamma(\Upsilon \to \mu^+ \mu^-)$.  To evaluate
$R_{\Upsilon}$ in the SM, we must include the three gluon final state
in the hadronic width as well as photon and $Z$-mediated dijet and
dimuon production~\cite{Appelquist:1974zd, Berger:1980ni}.  Since the
$Z$-mediated interference and contribution to the dimuon width is an
$\mathcal{O}(10^{-3})$ correction to the QED contribution, we treat
the dimuon width as a purely QED calculation for both the SM and
baryon-number calculation of $R_{\Upsilon}$.  Consequently, the
$|\Delta R_{\Upsilon}| = |R_{\Upsilon} - R_{\Upsilon}^{\text{SM}}|$
absolute difference cancels the three gluon contribution to the
hadronic width, so the modification of the dijet width provides the
leading sensitivity to the parameters $g_{_B}$ and $M_{Z'}$.  The
ARGUS constraint on the non-electromagnetic dijet decays of $\Upsilon$
gives $|\Delta R_{\Upsilon}| < 2.1$.

Thus, we calculate
\bear
\Delta R_{\Upsilon} = \frac{ \sum_{q} \left( \left|
\mathcal{M}_\gamma (q\bar{q}) + \mathcal{M}_{Z} (q\bar{q})
+ \mathcal{M}_{Z'} (q\bar{q}) \right|^2
- \left| \mathcal{M}_\gamma (q\bar{q}) + \mathcal{M}_{Z_\text{SM}} (q\bar{q})
\right|^2 \right)}{
|\mathcal{M}_\gamma (\mu^+ \mu^-) |^2} \ ,
\eear
where $\mathcal{M}_{Z} (q\bar{q})$ uses the modified $Z$ couplings,
Eq.~(\ref{eq:Zcouplings}).  In the limit that the $\Upsilon$ decay
products are massless and $Z'_B - Z_{\text{SM}}$ kinetic mixing is neglected, we obtain
\bear
\Delta R_{\Upsilon} = 
\frac{g_B^2 M_{\Upsilon}^2}{3 e^2 (M_{Z'}^2 - M_{\Upsilon}^2)} 
\left( 1 + 
\frac{g_B^2 M_{\Upsilon}^2}{4 e^2 (M_{Z'}^2 - M_{\Upsilon}^2)} + 
\frac{M_{\Upsilon}^2 (3 - 4 s_W^2)}{4 c_W^2 
(M_{Z}^2 - M_{\Upsilon}^2)} 
\right) \ .
\eear
If we also neglect the last term from $Z-Z'$ interference, this
expression agrees with Ref.~\cite{Carone:1994aa}.  The corresponding
constraint on $g_B$ is then
\bear
g_B < \sqrt{2} e \Bigg(
\sqrt{3 |\Delta R_\Upsilon| + 1} \left| 1 - \frac{M_{Z'}^2}{M_{\Upsilon}^2} \right|
+ \left( 1 - \frac{M_{Z'}^2}{M_{\Upsilon}^2} \right) \rule{0mm}{6mm} \Bigg)^{1/2} \ ,
\label{eq:gBfromUpsilon}
\eear
as shown in the low-mass gray-shaded region of
Figure~\ref{fig:bounds2}.  We have verified numerically that the
precise constraint on $g_B$ with finite final state masses and kinetic
mixing from Eq.~(\ref{eq:kappaZap}) gives a correction to
Eq.~(\ref{eq:gBfromUpsilon}) of less than 1\%.
Although our plot displays only masses above 5 GeV in Figure~\ref{fig:bounds2}, 
the constraint from $\Upsilon \to jj $ can be extended to lower $M_{Z'}$ values. For $M_{Z'} \sim 3$ GeV the constraint from $J/\psi$ decay \cite{Dobrescu:2014fca} is stringent, 
and below $\sim 1$ GeV additional experimental constraints become dominant \cite{Tulin:2014tya}.

In addition to limits from direct dijet resonance searches aimed at
the $Z'_B$ boson, and indirect constraints from $Z$--$Z'_B$ mixing and
$\Upsilon$ decays, we also have constraints on the anomalons, which
are required for self-consistency of the theory.  
One could introduce anomalons which replicate an entire generation of
SM fermions, but assign the new quarks $U(1)_B$ charge $-1$.  The new
fermions cancel the $[SU(2)_W]^2 \times U(1)_B$ and $[U(1)_Y]^2 \times
U(1)_B$ anomalies, which are linear in the $U(1)_B$ charges, and also
avoid generating new $U(1)_B$-gravity or $[U(1)_B]^3$
anomalies~\cite{FileviezPerez:2010gw}. Phenomenologically, however,
this solution is ruled out by the observed SM-like behavior of the
125~GeV Higgs boson, because the anomalons behave as a fourth
generation of chiral fermions, which exhibit non-decoupling behavior
in loop-induced Higgs processes.  While additional states can in
principle cancel these contributions~\cite{Kumar:2012ww}, the
non-decoupling nature of the anomalons in Higgs physics combined with
the direct production probes for new quarks essentially excludes this
solution. This discussion generalizes to any solution where the
anomalons are chiral under the SM gauge group.

A better option is to make the anomalons vectorlike under the SM gauge
group and chiral under $U(1)_B$.  Because the only mixed anomalies
from SM fields are $[SU(2)_W]^2 U(1)_B$ and $[U(1)_Y]^2 U(1)_B$, the
new anomalons do not have to carry color~\cite{Carena:2004xs,
  FileviezPerez:2011pt, Duerr:2013dza, Duerr:2013lka, Arnold:2013qja,
  Perez:2014qfa, Ohmer:2015lxa}, which significantly weakens their
direct production rates at the LHC.  Conversely, the anomalons do
carry electric charge and hence mediate a non-decoupling diphoton
partial width for the scalar associated with $U(1)_B$ breaking, which
we will explore in a further publication.

An extra feature for hadronic $Z'_B$ gauge bosons is the possible
1-loop vanishing of $Z$--$Z_B'$ mixing at the anomalon mass scale,
which amounts to a trace condition of all fermions charged under both
groups, Tr$(z_B Y) = 0$, with $z_B$ being the charges under
$U(1)_B$.  At energy scales below the anomalon masses, $Z$--$Z'_B$
mixing is reintroduced logarithmically.

Direct searches for the minimal model anomalons, in conjunction with a
theoretical upper limit on the Yukawa couplings, rule out an
additional region in the $Z'$ mass-coupling
plane~\cite{Dobrescu:2014fca}.  Recall that the anomalon masses are
generated by the Yukawa interactions (\ref{eqn:anomalons}) and the VEV
$v_{\phi} = 2 M_{Z'} / (3 g_{_B})$.  The perturbativity bound on the
anomalon Yukawa couplings is roughly given by $y_L$, $y_E$, $y_N
\lesssim 4 \pi/3$, so that the anomalon masses satisfy $m_f \lesssim
(8 \pi/9) M_{Z'} / g_{_B}$.  Thus, an experimental lower limit on the
anomalon masses translates into an upper limit on the gauge coupling
$g_{_B} < (8\pi/9) M_{Z'} / m_f$. The LEP constraint that the
anomalons must be heavier than about $90$~GeV rules out the
blue-shaded region in Figure~\ref{fig:bounds2}.

Stronger lower limits on the anomalon masses could be imposed by
searches at the LHC, but these are highly model dependent.  For small
mass splittings between the charged and neutral anomalons, their
collider phenomenology is similar to charged and neutral higgsino
collider searches at the LHC~\cite{ATLAS:2019lng, CMS:2023qhl,
  ATLAS:2023lkv}, which are searched for using multilepton
distributions and also metastable charged track signatures.  For
illustration, if an $m_f > 200$~GeV constraint is derived, then only
the region below the straight dashed line in Figure~\ref{fig:bounds2}
is allowed by the perturbativity bound discussed above.  It turns out,
however, that there are regions in the parameter space where the mass
limits on anomalons from the LHC are weaker than the one from LEP.

As an example, consider values for the Yukawa couplings in
(\ref{eqn:anomalons}) that, after $\phi$ is replaced by its VEV, give
the following anomalon mass terms: $m_f \bar L L + (m_f + \delta m_f)
\bar E E + (m_f - \delta m_f) \bar N N$.  Yukawa couplings of the
anomalons to the Higgs doublet of the type $y_1\bar L E H + y_2 \bar N
L H$ lead through mass mixing to the decays of the charged anomalons
to the neutral anomalons via off-shell $W$ bosons, where the final
state SM decay product is a charged pion (or a lepton and a neutrino, with much smaller branching fraction).
If $y_1, y_2 \approx \delta m_f/m_f$, then the decay width
of a charged anomalon to a charged pion and a neutral anomalon is of the order of  $(G_F^2 / \pi) 
 f_\pi^2 \, (\delta m)^3 \sqrt{1 - m_\pi^2 / (\delta m)^2}$,
where $G_F$ is the Fermi decay constant, and $f_\pi \approx 130.4$~MeV is the pion decay constant \cite{Thomas:1998wy}.    
In our case, for $\delta m = 1$~GeV the charged anomalon decay length is of the order of $10^{-2}$ mm in its
rest frame, and is prompt on collider length scales.
Additional hadronic or leptonic decays
become important as $\delta m$ increases, which also increases the
total decay width and shortens the charged anomalon lifetime.  While
Ref.~\cite{Thomas:1998wy} and subsequent collider phenomenology
studies~\cite{Arvanitaki:2012ps, Hall:2012zp, Bomark:2013nya,
  Han:2014kaa, Nagata:2014wma, Saito:2019rtg} and experimental
searches~\cite{ATLAS:2015wsk, ATLAS:2016tbt} have focused on the
metastable signature of charged winos and higgsinos, the very
difficult prompt decay signature into pions or charged leptons has
also been emphasized~\cite{Han:2014kaa}. 

Pair production of anomalons has the largest cross section when it is
mediated by an off-shell $W$ boson.  This cross section is $\sigma
(L^+ L^0) \approx 8$ pb for $m_f = 100$ GeV and $\delta m_f \ll m_f$
for $\sqrt{s} = 13$~TeV LHC \cite{Kumar:2015tna}, and is not large
enough to allow events with highly boosted pions.  Hence, the leading
collider searches focus on the leptonic signatures, where the heavier
neutral anomalon decays to the lighter neutral anomalon via an
off-shell $Z$ boson, giving a multilepton signature tested by ATLAS
and CMS~\cite{ATLAS:2019lng, CMS:2023qhl, ATLAS:2023lkv}.  Such
searches have significantly weakening sensitivity as $\delta m$
becomes smaller than 5~GeV, where only the LEP exclusion limit
survives when $\delta m \lesssim 1.5$~GeV.  Moreover, the anomalon
Drell-Yan cross section is about half of the higgsino Drell-Yan cross
section of 16.8 pb~\cite{Fuks:2012qx, Fuks:2013vua} for fully
degenerate $100$~GeV higgsinos at the 13~TeV LHC, since we only have
one $SU(2)$ doublet.  Thus, the leptonic signals from the anomalons
are also too weak to be seen by the LHC experiments thus far, and we
only adopt the LEP constraint in our study.

As discussed in~\cite{Preskill:1990fr} and emphasized recently
in~\cite{Cui:2017juz, Dror:2017ehi, Ismail:2017ulg, Dror:2017nsg,
  Ismail:2017fgq, Michaels:2020fzj}, the $Z$ boson can decay to $Z'$
and a photon via a Wess-Zumino-Witten interaction from the non-zero
anomaly induced by the non-decoupling effects of the anomalons as they
become heavy.  The full calculation of the partial width is found in
Ref.~\cite{Michaels:2020fzj}, where the physics of the anomalons and
the matching to the Wess-Zumino-Witten term is manifest.

From Ref.~\cite{Michaels:2020fzj}, the decay width of $Z \to Z'_B
\gamma$ is
\begin{align}
\Gamma(Z \to Z'_B \gamma) &= 
    \dfrac{\alpha_{\text{EM}} \alpha \alpha_B}{384 \pi^2 c_W^2} \dfrac{m_Z'^2}{m_{Z}} \left( 1 - \dfrac{m_{Z'}^4}{m_Z^4} \right) \nonumber \\
    &\Bigg| 
-\sum\limits_{f \in \text{ SM }} T_3(f) Q_f^e
    \left[ \dfrac{m_Z^2}{m_Z^2 - m_{Z'}^2} \left( B_0(m_Z^2, m_f) - B_0(m_{Z'}^2, m_f) \right) + 2 m_f^2 C_0(m_f) \right]
    \nonumber \\
    &+ 3 \left( \dfrac{m_Z^2}{m_Z^2 - m_{Z'}^2} \left( B_0 (m_Z^2, M) - B_0 (m_{Z'}^2, M)  \right) + 2 M^2 \dfrac{m_Z^2}{m_{Z'}^2} C_0 (M) \right)
\Bigg|^2 \ ,
\label{eqn:ZZbgamma}
\end{align}
where $T_3(f) = +1$ for up-type quarks and $-1$ for down-type quarks,
$M$ is the mass of the anomalons and assumed to arise only from
$U(1)_B$ breaking, and $C_0$ and $B_0$ are the usual Passarino-Veltman
3-pt.~and 2-pt.~scalar integrals, following the conventions of
\textsc{Package-X}~\cite{Patel:2015tea, Passarino:1978jh},
\begin{align}
B_0 (m_V^2, m) \equiv B_0 (m_V^2, m, m) \ , \quad
C_0 (m) \equiv C_0 (0, m_Z^2, m_Z'^2, m, m, m) \ .
\end{align}
We can construct an approximate expression for
Eq.~(\ref{eqn:ZZbgamma}) by taking the first five SM quarks to be
massless while the top quark and anomalons are taken to infinity.
Note this expression is still only valid when the anomalon masses are
solely generated from $U(1)_B$ breaking.  The approximate partial width is then
\begin{align}
\Gamma(Z \to Z'_B \gamma) &\approx
\dfrac{\alpha_{\text{EM}} \alpha \alpha_B}{384 \pi^2 c_W^2} \frac{m_{Z'}^2}{m_Z} 
\left( 1 - \frac{m_{Z'}^4}{m_Z^4} \right) 
\left| \frac{3 m_Z^2}{m_{Z'}^2} - \frac{2}{3} 
- \frac{7}{3} \frac{m_Z^2}{m_Z^2 - m_{Z'}^2} 
\log \left( \frac{m_Z^2}{m_{Z'}^2} \right) \right|^2 \ .
\end{align}
The pink region in Figure~\ref{fig:bounds2} shows the
limit calculated using Eq.~(\ref{eqn:ZZbgamma}) from the search of the
L3 experiment for the exotic $Z$ decay, $Z \to Z' \gamma$, $Z' \to
jj$~\cite{Adriani:1992zm}, also taken from
Ref.~\cite{Michaels:2020fzj}, where anomalon masses are fixed with
Yukawa couplings $4\pi/3$ and arise solely from $U(1)_B$ breaking.
The exotic decay constraint by L3 is competitive with the 35.9
fb$^{-1}$ ISR $\gamma$ search by CMS~\cite{CMS:2019xai}, although the
indirect bound for charged anomalons still provides the dominant
constraint~\cite{Dobrescu:2014fca}.


\section{Conclusions}
\setcounter{equation}{0}
\label{sec:conclusions}

We have analyzed the current state of the experimental collider
searches for dijet resonances, and compared them with the electroweak
constraints on a $Z'$ boson.  Notably, the LHC experiments now provide
the leading constraints not only at masses of several TeV, but also on
electroweak scale dijet resonances, thanks to the advent of new
trigger pathways and advanced data reconstruction methods.

In addition, the ATLAS and CMS experiments are also placing direct
dijet bounds on resonances below 100~GeV, where legacy measurements
from LEP experiments, constraints from $\Upsilon$ meson measurements,
and indirect limits on charged anomalons compete for the strongest
sensitivity.  We have emphasized that in gauged $U(1)_B$ models where
the fermion sector obeys the orthogonality condition in
Eq.~(\ref{eq:trace}), the kinetic mixing between SM and $Z'$ gauge
bosons is finite and only logarithmically sensitive to the anomalon
masses.  Moverover, the contribution of the anomalons to the exotic
decay $Z \to Z'_B \gamma$ also follows non-decoupling behavior of
chiral fermions, reducing the sensitivity on their mass scale.  Thus,
the coupling-mass plot, which is an insightful way of presenting the
collider limits on dijet resonances, also allows a meaningful
comparison with the low-energy data.  Our summary of collider
constraints, shown in Figure~\ref{fig:bounds2}, also includes the
competing bounds from modifications to the properties of the $Z$ and
$\Upsilon$ due to the $Z'_B$, as well as the indirect limits from the
Yukawa couplings of the anomalons.

We have also emphasized that, like the SM, the underlying
chiral structure of the $U(1)_B$ symmetry is characterized by a single
VEV, and hence the $Z'_B$ and anomalon masses cannot be
arbitrarily decoupled from each other without violating perturbative
unitarity.  In the coupling versus mass plane, the improving
constraints continue to probe higher scales of $U(1)$ symmetry
breaking, as evident from the diagonal lines corresponding to constant
$m_f$ anomalon masses in Figure~\ref{fig:bounds2}.  The possible
sensitivity improvements from collider searches for anomalons as well
as signals of the symmetry breaking sector 
(see, {\it e.g.}~Ref.~\cite{Duerr:2017whl}) are left for future work.

\bigskip\bigskip\bigskip 

\noindent
{\it Acknowledgments: } FY is supported by the Cluster of Excellence
PRISMA$^+$, ``Precision Physics, Fundamental Interactions and
Structure of Matter'' (EXC 2118/1) within the German Excellence
Strategy (project ID 390831469).  Fermilab is administered by Fermi
Research Alliance, LLC under Contract No. DE-AC02-07CH11359 with the
U.S. Department of Energy, Office of Science, Office of High Energy
Physics.

  
\section*{Appendix: \ $Z'-Z$ mixing}  \addcontentsline{toc}{section}{Appendix: \ $Z'-Z$ mixing}  
\label{sec:appendix}     
\renewcommand{\theequation}{A.\arabic{equation}}
\setcounter{equation}{0}

In this Appendix we compute the kinetic and mass mixings of the gauge
eigenstate $Z'_{\rm g.e.}$ boson with the $Z_{\rm SM}$ boson, for
general couplings ($z^f \, $) of $Z'_{\rm g.e.}$ to the fermions that
satisfy the orthogonality relation Tr$(Y z) = 0$, where $Y$ is the
hypercharge.

The 1-loop amplitude for $Z'_{\rm g.e.}-Z_{\rm SM}$ mixing induced by
fermions is given by $\epsilon_\mu \epsilon_\nu {\cal A}^{\mu\nu}$,
where $ \epsilon_\mu, \epsilon_\nu$ are the polarization vectors of
the two gauge bosons, and
\bear
{\cal A}^{\mu\nu} & \!\! = \!\! &  i \, \frac{g_{_B} g}{ c_W} \mu^{4-D}  \int \!\! \frac{d^D k}{(2\pi)^D}  \sum_f
\frac{
N_f}  {\left[ (p+k)^2 - m_f^2\right]  \left( k^2 - m_f^2 \right) }
\left\{  
m_f^2 \left( g_L^f  z_R^f + g_R^f  z_L^f  \right) g^{\mu\nu}     \right.
\nonumber \\ [2mm]
& &  +  \left.  
\left( g_L^f  z_L^f + g_R^f  z_R^f  \right)\left[  p^\mu k^\nu + k^\mu p^\nu + 2 k^\mu k^\nu - g^{\mu\nu} (p+k)\cdot k  \rule{0mm}{4mm} \right]
\right\}    ~~.
\eear
Here $p$ is the 4-momentum of the $Z'_{\rm g.e.}$ and $Z_{\rm SM}$
bosons, and we used dimensional regularization with $D = 4 -
\varepsilon$ and a scale $\mu$.  The above sum is over the fermions
$f$, which have a color factor $N_f$.  Their right- and left-handed
components ($f_R$ and $f_L$) carry $Z'_{\rm g.e.}$ charges $z^f_R$ and
$z^f_L$, respectively, and also have couplings ($g^f_R$ and $g^f_L$)
to the $Z_{\rm SM}$ boson. After combining the denominators, we get
\bear  \hspace*{-3cm}
{\cal A}^{\mu\nu} &  \!\!  =  \!\!  &  i \, \frac{ g_{_B} g}{ c_W} \;  \int_0^1  \!  dx \,  \sum_f  N_f   \left\{  \rule{0mm}{6mm}     m_f^2  \left( g_L^f  z_R^f + g_R^f  z_L^f  \right) g^{\mu\nu}  I_0^f    \right.
\nonumber \\ [2mm]
& &  + \left. 
\left( g_L^f  z_L^f + g_R^f  z_R^f  \right)  \left[   g^{\mu\nu} I_1^f  + x (1-x)   \left( g^{\mu\nu} p^2 - 2 p^\mu p^\nu  \right)   I_0^f   \right]  \rule{0mm}{6mm}   \right\}   ~~,
\label{eq:Amunu}
\eear
where  $ I_0^f$ and  $ I_1^f$ are the following integrals:
\be 
\left\{ I_0^f  \, , \,  I_1^f  \right\} = \mu^{4-D}    \int \!\! \frac{d^D k}{(2\pi)^D}   \frac{1 }{ \left[ k^2 - m_f^2 + x (1-x) p^2 \right]^2  }    \left\{ 1 \, , \,  k^2   \left( \frac{2}{D } - 1 \right)  \right\}   ~~.
\ee

The orthogonality relation Tr$(Y z) = 0$ implies that the fermion
charges satisfy
\be
\sum_f  N_f \left( g_L^f  z_L^f + g_R^f  z_R^f  \right) = 0  ~~.
\label{eq:sumRule}
\ee 
Consequently, the apparent quadratic divergence of $ I_1^f $ in the
$D=4$ limit vanishes after the sum over fermions is performed. The
usual $\varepsilon$ expansion and the $\overline{\rm MS}$ scheme lead
to the following expressions for the integrals:

\bear
&& I_0^f  = \frac{- i}{ (4 \pi)^2 }  \,  \ln \left( \frac{m_f^2}{\mu^2} - x (1-x) \frac{p^2}{\mu^2}  - i\epsilon_0 \right)  ~~,
   \nonumber \\ [-2mm]
  \label{eq:integralsI}
 \\ [-1mm]
&& I_1^f  = -  \left( m_f^2 - x (1-x) p^2   \rule{0mm}{4mm}    \right)  I_0^f  ~~~,
\nonumber
\eear
where $ i\epsilon_0$ is the prescription for the complex logarithm
when $m_f^2 < x (1-x) p^2$.

The real part of the $Z'_{\rm g.e.}-Z_{\rm SM}$ mixing amplitude
contains two pieces, as shown in Eq.~(\ref{eq:Amunuform}): a kinetic
mixing (with dimensionless coefficient $\kappa_Z$), and a mass mixing
$\Delta M^2_{Z'Z}$ (the off-diagonal entry in the mass-squared matrix
for the two gauge bosons).  From Eq.~(\ref{eq:Amunu}) and the second
Eq.~(\ref{eq:integralsI}) follows that
\bear  \hspace*{-3cm}
 \kappa_Z   &  \!  =  \!  &  2  \, \frac{g_{_B} g}{ c_W} \; \sum_f  N_f  \left( g_L^f  z_L^f + g_R^f  z_R^f  \right)  \, {\rm Re} \; i \! \int_0^1  \!  dx \,  x (1-x)  I_0^f    ~~~,
  \nonumber \\ [-3mm]
  \label{eq:kappa}
 \\ [-2mm]
  \Delta M^2_{Z'Z}  &  \!  =  \!  & - \, \frac{g_{_B} g}{ c_W} \; \sum_f  N_f \,  m_f^2   \left( g_L^f  - g_R^f  \right)   \left( z_L^f -  z_R^f  \right)   \, {\rm Re} \; i \!  \int_0^1  \!  dx \,   I_0^f     ~~~.
    \nonumber
 \eear
Integrating over $x$, we find
\be   
 \int_0^1  \!  dx \,  x (1-x) \, I_0^f  = \frac{- i}{ 6  (4 \pi)^2 }  \left[   \ln \! \left( \frac{p^2}{\mu^2} \right)  - \frac{5}{3}  + {\cal F}(m_f^2/p^2) 
+  i \pi \, {\cal G}(m_f^2/p^2)  \right]  ~.
\label{eq:xI0}
 \ee
The functions introduced here are ${\cal G}(y) = \theta ( 1 - 4 y) \,
\left( 1 + 2y \right) \sqrt{ 1 - 4 y } \, $, where $ \theta $ is the
step function, and
\be \hspace*{-0.4cm}
{\cal F}(y) =  \ln  y - 4 y +   (1 + 2 y) \, \left| 4 y - 1 \rule{0mm}{3.8mm}  \right|^{1/2}  \! \times \!
 \left\{ \ba{c}  
 \displaystyle   \ln \! \left( \frac{  \displaystyle 1+  \! \sqrt{  1 - 4 y  }  }{2y} -1 \! \right) 
  \; \,  {\rm for} \;\, y \leq \frac{1}{4}     ~~,
\\ [5mm]  
 \displaystyle  2  \,  \arctan \! \left[ (4y -1)^{-1/2}  \rule{0mm}{3.9mm} \right]    \;\,  {\rm for} \;\, y > \frac{1}{4}    ~~.
\ea \right.
\label{eq:Fyfull}
\ee

We emphasize that the sum over fermion loops removes not only the quadratic divergence from $ I_1^f $, but also 
the logarithmically divergent part of each fermion loop, which is shown in (\ref{eq:xI0}).
Thus, the coefficient for kinetic mixing of any $Z'_{\rm g.e.}$ and $Z_{\rm SM}$ of 4-momentum $p$
is {\it finite}, and can be written as the following sum over the fermion loops:
\bear
 \kappa_Z   &  \!  \simeq \!  & \!  \frac{ g_{_B} g}{ 48 \pi^2 c_W } \,  \sum_{f} \,  N_f  \left( g_L^f  z_L^f + g_R^f  z_R^f  \right)  
  {\cal F}(m_f^2/p^2)   ~~.
\label{eq:kappaZ} 
\eear
This result is used in Section~\ref{sec:mixing}. The expansion of the
function ${\cal F}(y)$ for $y \ll 1/4$ is
\be
 {\cal F}(y)  \simeq  -6  y  +  6 y^2 \ln y  +    O \!  \left( y^2  \right)  ~~,
 \label{eq:Fylow}
\ee
while for $y \gg 1/4$
\be
 {\cal F}(y)  \simeq   \ln  y  + \frac{5}{3} - \frac{1}{5 y} + O \! \left((4y)^{-2}  \rule{0mm}{3.86mm}  \right)  ~~.
\ee
   
  
\providecommand{\href}[2]{#2}\begingroup\raggedright

\vfil
\end{document}